\documentclass{aa}  
\usepackage{natbib}
\usepackage{graphicx}
\usepackage{subfig}
\usepackage{txfonts}
\usepackage{caption}
\usepackage{hyperref}
\hypersetup{
    colorlinks=true,
    linkcolor=red,
    filecolor=cyan,
    citecolor=blue,
    urlcolor=magenta
     }

\begin{document}

   \title{Interactions among intermediate redshift galaxies}

   \subtitle{The case of SDSSJ134420.86+663717.8\thanks{This paper uses data taken with the MODS spectrographs built with funding from NSF grant AST-9987045 and the NSF Telescope System Instrumentation Program (TSIP), with additional funds from the Ohio Board of Regents and the Ohio State University Office of Research.}}

   \author{Persis Misquitta
          \inst{1}\thanks{ \email{misquitta@ph1.uni-koeln.de}},
          Micah Bowles\inst{1}, Andreas Eckart\inst{1}\fnmsep \inst{2}, 
          Madeleine Yttergren\inst{1}\fnmsep \inst{2}\thanks{Member of the International Max Planck Research School (IMPRS) for Astronomy and Astrophysics at the Universities of Bonn and Cologne.}, 
          Gerold Busch\inst{1}, Monica Valencia-S.\inst{1}\fnmsep\inst{3}, Nastaran Fazeli\inst{1}
          }

   \institute{I. Physikalisches Institut, Universität zu Köln, Zülpicher Str. 77, 50939, Köln, Germany\\
   \and
   Max-Planck Institut für Radioastronomie (MPIfR), Auf dem Hügel 69, 53121 Bonn, Germany\\
   \and
   Regional Computer Center (RRZK), Universität zu Köln, Weyertal 121, 50931, Köln, Germany\\
   }

   \date{Received October 28, 2019; accepted May 9, 2020}

  \abstract
  {We present the properties of the central supermassive black holes and the host galaxies of the interacting object SDSSJ134420.86+663717.8. We obtained optical long slit spectroscopy data from the Large Binocular Telescope (LBT) using the Multi Object Double Spectrograph (MODS). Analysing the spectra revealed several strong broad and narrow emission lines of ionised gas in the nuclear region of one galaxy, whereas only narrow emission lines were visible for the second galaxy. The optical spectra were used to plot diagnostic diagrams, deduce rotation curves of the two galaxies, and calculate the masses of the central supermassive black holes. We find that the galaxy with broad emission line features has Seyfert~1 properties, while the galaxy with only narrow emission line features seems to be star-forming in nature. Furthermore, we find that the masses of the central supermassive black holes are almost equal at a few times $10^7~M_\odot$. Additionally, we present a simple N-body simulation to shed some light on the initial conditions of the progenitor galaxies. We find that for an almost orthogonal approach of the two interacting galaxies, the model resembles the optical image of the system.}

   \titlerunning{Interaction of intermediate redshift galaxies}
   \authorrunning{P.~Misquitta et al.}
   \maketitle
   
%
%________________________________________________________________

\section{Introduction}
  Interactions play an important role in the evolution of galaxies. It is generally accepted that smaller galaxies in the past merged to form larger 
  galaxies \citep{White, Kauffmanna, Cole, Casteels}. The gas content in these galaxies was affected as a result of this process. Gas-rich galaxies interacted with one another, causing the gas and dust content to get funneled through to the center, leading to bursts of star formation and enabling the black holes in the centers of galaxies to start accreting material \citep{Barnesc, Barnes, Mihosa, Mihosb, DiMatteo, Hopkinsa, Hopkinsb, Cox, Lotza, Lotzb, Lotzc, DiMatteoa, Ellison, Torrey, Casteels}. Accretion onto the central super massive black hole (SMBH) and the starburst phase leads to the depletion of gas in the progenitor galaxies and the merger remnants end up gas-poor \citep{Springel, Johansson, Hopkins}. 
  
  Early surveys and catalogs by \citet{V-V, Arp, Zwicky} list many galaxies as distorted or irregular. \citet{Pfleiderer, Pfleiderera, Yabushita, Toomre, Clutton-Brocka, Clutton-Brockb, Wright, Eneev, Hernquist} showed that the distortions could be caused by the influence of the gravitational potential of one galaxy on the gas and dust content of the other galaxy using simple models and three-body interactions. Over the years, many attempts have been made to study the merger rate over cosmic time. For example: \citet{Mantha} analysed a sample of 9800 massive ($M_\mathrm{{stellar}}\geq2\times10^{10}~M_\odot$) galaxies spanning $0<z<3$, and found that the merger rate fraction increases from $z\approx0$ to $0.5<z<1$ before decreasing at $2.5<z<3$, indicating a turnover at $z\approx1$. \citet{Hopkinsd} provides a good review of the different models used to study merger rates theoretically. 
  
  Another interesting point to consider while studying interacting galaxies is the influence they have on triggering active galactic nuclei (AGN). While traditionally, the theoretical view has been that mergers are the predominant mechanism responsible for black hole growth and AGN activity, some theoretical studies predict the requirement of other processes along with merging to drive nuclear activity \citep{Fanidakis, Hirschmann}. \citet{Steinborn} conclude that while mergers increase the probability for nuclear activity, the role of mergers causing nuclear activity in the overall AGN population is still minor (<20~\%). This is also the case with observational studies, with some of them finding that mergers play a minor role in triggering luminous AGN \citep{Villforth, Hewlett}, while other studies find that mergers are the dominant mechanism for driving AGN activity \citep{Urrutia, Hopkins2009, Treister, Glikman, Fan}. 
  
  The highly energetic phenomena known as AGN emit brightly over the entire electromagnetic spectrum. They can be broadly classified as Type~1 and Type~2 AGN based on the respective presence or absence of broad emission lines in their spectra. \citet{Antonucci} put forth an AGN unification model, wherein Type~1 and Type~2 AGN are the same type of object, with the difference appearing relative to the orientation of the observer. If the unification scheme holds true, there should be no significant difference in the external environment of Type~1 and Type~2 AGN. However, \citet{Dultzin, Koulouridis, Jiang} found more Type~2 galaxies than Type~1 galaxies in close galaxy pairs, leading to the suggestion that instead of a unified AGN model, galaxy interactions could result in evolution of galaxies from star-forming to Type~2 and finally to Type~1. \citet{Gordon} find that upto a pair separation of $50~\mathrm{kpc~h^{-1}}$, there is no difference in the AGN found in pairs based on the AGN type, with an excess of Type~2 AGN at separations lesser than $20~\mathrm{kpc~h^{-1}}$. They suggest that either the unified model breaks down in the case of close gravitational interactions or that the interaction between galaxies drives more dust towards the nucleus, causing the obscuration to be more probable. Additionally, \citet{Ricci2017} find that the fraction of Compton-thick AGN in late merger galaxies is higher compared to the local hard X-ray selected AGN. They conclude that late stage mergers of galaxies suffer greater obscuration as material is most efficiently funneled to the inner parsecs from the outer scales and the classical AGN unification model cannot sufficiently explain this phenomenon. 
  
  The lifetimes of interactions among galaxies are measured in the order of $\mathrm{10^{8}}$ years \citep{Struck}. It is, therefore, impossible for us to study a particular case in the entirety of the process. Every case available to us is a cosmic 'snapshot' of a particular phase in the long series of interactive events, and as such, yields information and helps to better our understanding of galaxy mergers. Studying different pairs of galaxies at different stages in their interaction aids in putting together a coherent picture of the series of events. Another important tool that helps in increasing our knowledge is simulations of galaxy interactions. Observations help make the simulations more detailed while simulations help to interpret the observations.
  
\begin{figure}[h]
     \centering
     \includegraphics[width=0.45\textwidth]{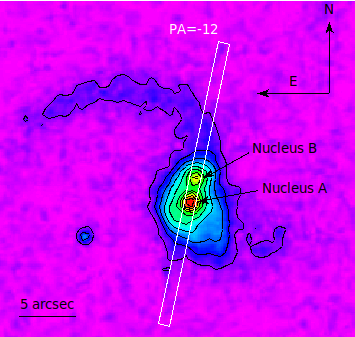}
     \caption{SDSS image of the galaxy pair SDSSJ134420.86+663717.8. North is up and left is East. The central bulge hosts the two nuclei: nucleus A, which is at the bottom of the bulge, and nucleus B, at the top of the bulge. Clearly visible is the tidal arm to the north-east of nucleus B, while the tidal arm to the south-west of nucleus A appears quite foreshortened. The position angle in degrees (measured from the North to the East) and the 1~arcsec slit used for long-slit spectroscopy with LBT/MODS are illustrated. Credit: SDSS Skyserver Data Release 15. }%
     \label{fig:sdss}
   \end{figure}

  SDSSJ134420.86+663717.8 is a pair of interacting galaxies that is located at an intermediate redshift of approximately $z\sim0.128$\footnote{\url{http://skyserver.sdss.org/dr15/en/tools/explore/Summary.aspx?id=1237651273514745920}}. The pair derives its name from the Sloan Digital Sky Survey catalogue number. Previously, they were cataloged as a clumpy source by several surveys like the Infrared Astronomical Satellite (IRAS), the Wide-field Infrared Survey Explorer (WISE), the Two Micron All Sky Survey (2MASS), the ROentgen SATellite (ROSAT) survey, and the Palomar Schmidt 48-inch telescope survey. Table \ref{table:lit_values} presents the literature values of luminosity and photometry for SDSSJ134420.86+663717.8. The Sloan Digital Sky Survey (SDSS) provides a significantly improved picture of the galaxies. The two nuclei are well resolved and tidal distortions are clearly visible (see Figure \ref{fig:sdss}). The orientation and size of the tidal tails hints at an orthogonal interaction between the progenitor galaxies. The SDSS website lists some information about the larger of the two nuclei, which we shall call nucleus A, but provides no information for nucleus B. Nucleus A is classified as a quasi-stellar object (QSO) by SDSS \citep{Schneider2003}, as a Seyfert 1 galaxy by \citet{Zuther}, and as a Seyfert 1.5 galaxy by \citet{Paturel} and \citet{Moran}. \citet{Zuther} also note that SDSSJ134420.87+663717.6 is an interacting galaxy and the pair shows 2.70$\pm$0.40~mJy peak flux at 20~cm (taken from the NVSS 20~cm survey), $L_\mathrm{{FIR}}\approx5\times10^{11}~L_\odot$, and $\mathrm{\log} M_\mathrm{{BH}}=7.6$ (taken from \citet{Greene2007}). However, they could not detect any compact flux density at 18~cm with the Multi-Element Radio Linked Interferometer Network (MERLIN). They conclude that the radio emission was likely dominated by extended star formation since it is an infrared (IR) luminous galaxy. \citet{Moran} provide values for the flux and luminosity of SDSSJ134420.86+663717.8 (or IRASF13429+6652, after its IRAS catalogue name) in the X-ray and far infrared (FIR) regions to be $F_\mathrm{X}=4.88\times10^{-13}~\mathrm{erg~s^{-1}~cm^{-2}}$, $L_\mathrm{X}=3.87\times10^{43}~\mathrm{erg~s^{-1}}$, $F_\mathrm{{FIR}}=2.18\times10^{-11}~\mathrm{erg~s^{-1}~cm^{-2}}$, and $L_\mathrm{{FIR}}=1.73\times10^{45}~\mathrm{erg~s^{-1}}$, respectively. The X-ray luminosity was calculated using ROSAT observations over the energy range between 0.1-2.4 keV. 
  
  As an on-going merger of two intermediate redshift, almost orthogonal galaxies, SDSSJ134420.86+663717.8 is a subject of interest as a case study for orthogonal interactions of galaxies. This work focuses on studying the spectra of the two nuclei in the optical wavelength regime and analysing this data. In addition, we use a simple N-body simulation to infer the initial conditions of the host galaxies before the interaction, based on their morphology.  
  
  \begin{table*}[h!]
   \caption{Photometry Values for SDSSJ134420.86+663717.8}
   \smallskip
   \centering
   \scalebox{0.9}{
   \begin{tabular}{c c c c}
   \hline \hline Spectral Region & Band & Apparent Mag or Flux & Reference\\
   \hline Ultraviolet & NUV (GALEX) AB & 17.0747$\pm$0.0326211 mag & 1\\
   Visual & i (SDSS Petrosian) AB & 15.546$\pm$0.012 asinh mag & 2\\
   Near-Infrared & K$_s$ & 12.747$\pm$0.079 mag & 3\\
   Mid-Infrared & 12 microns (IRAS) & <5.397e-02 Jy & 3\\
   Far-Infrared & 60 microns (IRAS) & 4.074e-01 $\pm$9 \% Jy & 4\\
   Radio & 1.4 GHz & 3.2$\pm$0.5 mJy & 5\\
   \hline
   \end{tabular}}
   \label{table:lit_values}
   \tablebib{(1)~\citet{Seibert}; (2)~\citet{Schneider2002}; (3)~\citet{Cutri}; (4)~\citet{Moshir}; (5)~\citet{Condon} }
   \tablefoot{Credit:NED}
   \end{table*}
   
  This paper is organised as follows. We present details about the observations and data reduction in Section~\ref{section2}. In Section~\ref{section3}, we detail the process of spectral extraction and present the optical spectra of the two nuclei. Further on, we mention the results from emission line fits. We analyse the spectra in detail and list some of the properties of the galaxies in Section~\ref{4} and present an N-body simulation to model a SDSS~J1334420.86+663717.8 look-alike in Section~\ref{4.5}. Finally, we summarise the paper and provide some concluding remarks in Section~\ref{section5}.
%__________________________________________________________________

\section{Observations and data reduction}\label{section2}

 SDSSJ134420.86+663717.8 was observed using the Multi-Object Double Spectrograph (MODS) at the Large Binocular Telescope (LBT)\footnote{The LBT is an international collaboration among institutions in the United States, Italy, and Germany. LBT Corporation partners are: The University of Arizona on behalf of the Arizona Board of Regents; Instituto Nazionale di Astrofisica, Italy; LBT Beteiligungsgesellschaft, Germany, representing the Max-Planck Society, The Leibniz Institute for Astrophysics Potsdam, and Heidelberg University; The Ohio State University, and The Research Corporation, on behalf of The University of Notre Dame, University of Minnesota, and University of Virginia.} located at an altitude of approximately 3,200~m on Mount Graham, United States \citep{Pogge}. The LBT has a binocular design and consists of two $\mathrm{8.4~m}$ wide mirrors mounted side-by-side with a centre-to-centre distance of 14.4~m, such that it has a combined collecting area of an 11.8~m telescope. MODS are a pair of seeing-limited low to medium resolution, two-channel, identical spectrographs: MODS1 and MODS2. They are mounted at the direct Gregorian foci of the twin LBT mirrors. A dichroic splits the light into red and blue optimised spectrograph channels, such that the data corresponding to an object is obtained from four different channels: MODS1~B, MODS1~R, MODS2~B, and MODS2~R. MODS can be used for imaging, long-slit and multi-object spectroscopy. It has a field-of-view of $\mathrm{6^\prime\times6^\prime}$. The wavelength range for MODS is from 3000~\AA~to 10000~\AA. We used MODS in long slit spectroscopy mode to obtain the data on May 13, 2016. The exposure time for each spectrum was 600 seconds, and the object was observed thrice, with a total exposure time of 1800 seconds. In total, we had twelve spectra, corresponding to three spectra each from MODS1~B, MODS1~R, MODS2~B and MODS2~R, respectively. The slit width was 1~arcsec, with spectral resolution of $\mathrm{150~kms^{-1}}$, and a spatial resolution of 0.627~arcsec.
 
 The observed frames were bias corrected and combined using the MODS CCD Reduction pipeline \citep{richard_pogge}. Wavelength and flux calibration, and background subtraction was done using Pyraf routines developed by us, yielding two-dimensional spectra that were used to extract one-dimensional spectra. Hg, Ne, Ar, Xe, and Kr lamps were used for wavelength calibration while data from a standard star, BD+33d2642, was used for flux calibration. BD+33d2642 is a spectrophotometric standard star of spectral type B2IV with an established spectrum \citep{Oke}. It was observed using MODS during the same night as the target, SDSS J134420.86+663717.8. The standard star data underwent the same procedures as the science data in that it was bias corrected, combined and wavelength calibrated using the same methods. The data from MODS1 and MODS2 was reduced separately but was combined after reduction to improve the signal-to-noise ratio ($\mathrm{S/N}$). 
 
 \section{Emission line fits and results}\label{section3}
 We extracted one-dimensional spectra of nuclei A and B from the reduced two-dimensional spectra. This was done by choosing an aperture of approximately 1~arcsec on the spatial axis for each nucleus and then extracting along the wavelength axis. Additionally, we chose an aperture of approximately 8~arcsec and extracted spectra along each pixel row (1 pixel = 0.123 arcsec for MODS Red and 1 pixel = 0.120 arcsec for MODS Blue) in order to study the intensity distribution of the two nuclei. The continuum was fit with a line and all broad emission lines were fit using multiple Gaussian functions, while the narrow emission lines were fit using single Gaussian functions using the Levenberg-Markwardt algorithm, non-linear least-squares minimisation and curve-fitting for Python (LMFIT) \citep{Markwardt, Newville}.
 
 \subsection{Optical spectrum of Nucleus A} 
 \begin{figure*}[h]
     \centering
     \includegraphics[width=1\textwidth]{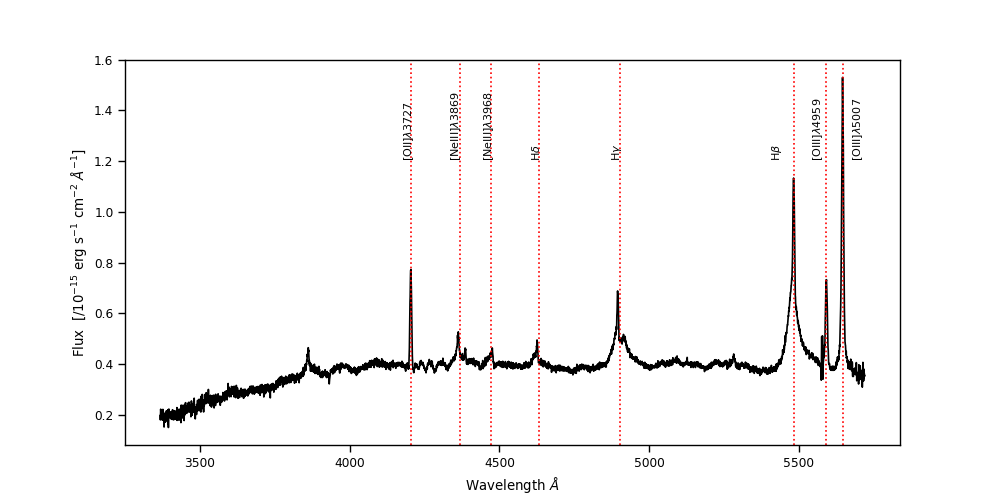}
     \caption{One-dimensional observed optical spectrum of SDSSJ134420.86+663717.8 nucleus A in the wavelength range $3500-5700~\AA$. The emission lines observed have been marked by vertical red dotted lines and named. The aperture size for spectral extraction is 1 arcsec. The wavelength axis depicts the observed wavelength.}%
     \label{fig:nuc_a_b}
   \end{figure*}
   
   \begin{figure*}[h]
     \centering
     \includegraphics[width=1\textwidth]{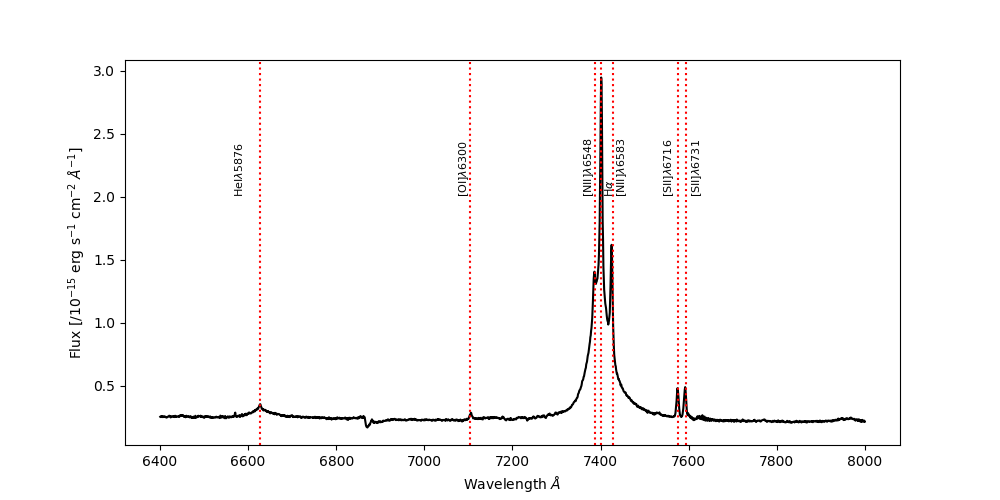}
     \caption{One-dimensional observed optical spectrum of SDSSJ134420.86+663717.8 nucleus A in the wavelength range $6400-8000~\AA$. The emission lines observed have been marked by vertical red dotted lines and named. The aperture size for spectral extraction is 1 arcsec. The wavelength axis depicts the observed wavelength.}%
     \label{fig:nuc_a_r}
   \end{figure*}
   
\begin{figure*}[h]
     \centering
     \includegraphics[width=1\textwidth]{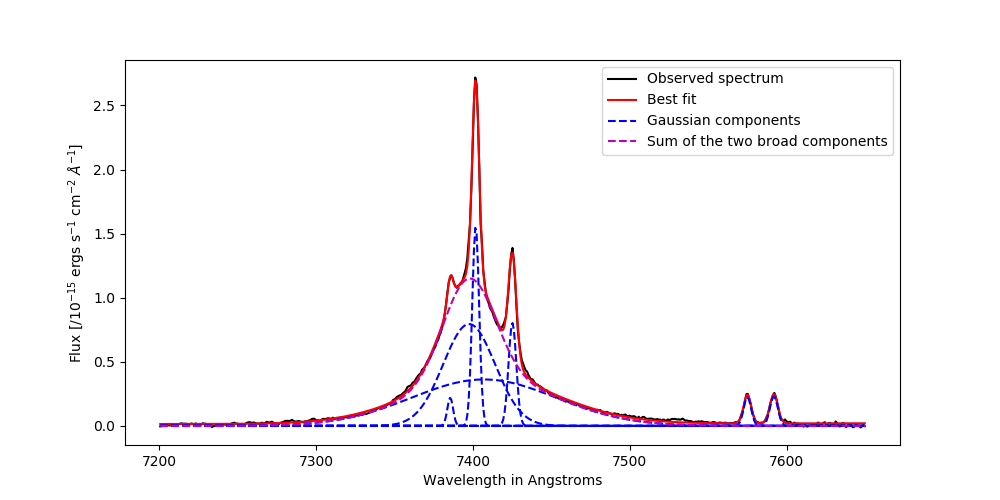}
     \caption{One-dimensional observed optical spectrum in the wavelength range 7200-7650~$\AA$ of SDSSJ134420.86+663717.8 nucleus A overlaid with the Gaussian fits. Two broad components along with three narrow components are fit to the $\mathrm{H\alpha} + \ion{N}{II}$ complex. The black line represents the observed spectrum, the red line is the best fit, the dotted blue lines are the individual Gaussian components, and the dotted magenta line is the sum of the two broad components. The wavelength axis depicts the observed wavelength.}%
     \label{fig:gauss}
   \end{figure*}

   Figures \ref{fig:nuc_a_b} and \ref{fig:nuc_a_r} show the optical spectrum of nucleus A. The continuum emission is overlaid by several emission lines. Stellar absorption features are not detected. Broad \ion{H}{I} emission features are clearly visible. The break in the wavelength axis reproduces the split between the blue and red beams due to the dichroic. Gaussian fits to the emission lines helped in determining the full width at half maximum (FWHM) of the various emission lines (see Figure~\ref{fig:gauss} for an example of the spectrum fit with Gaussian functions). In order to determine the number of Gaussian components to be fit, we started with, for example, three narrow Gaussian components and one broad component for the case of the H$\mathrm{{\alpha}+[\ion{N}{II}]}$ complex. We used visual inspection as well as the $\chi^2$ values of the Gaussian fitting module\footnote{The $\chi^2$ value for only one broad component is 1.19, whereas we find a  $\chi^2$ value of 0.32 for two broad components.} and determined the number that was the best fit to the data, which in our case was two broad $\mathrm{H\alpha}$ components and three narrow components corresponding to $\mathrm{[\ion{N}{II}]\lambda6548, H\alpha,}$ and $ \mathrm{ [\ion{N}{II}]\lambda6583}$, respectively. The choice was obvious: an additional broad component is required to fit the extended $\mathrm{[\ion{N}{II}]\lambda6548, H\alpha,}$ and $\mathrm{[\ion{N}{II}]\lambda6583}$
   wing towards longer wavelengths. The three narrow components are needed to fit the three narrow line features related to the $\mathrm{[\ion{N}{II}]\lambda6548, H\alpha,}$ and $\mathrm{ [\ion{N}{II}]\lambda6583}$ lines,
   respectively. Hence, our choice comprises only the most necessary components. Table \ref{table:flux_a} lists the different emission lines observed in the spectrum, their flux values and FWHM. The average redshift of nucleus A, as calculated using the central positions of various emission lines is $\mathrm{0.1279\pm0.0002 (\sim 38370~km~s^{-1})}$. The velocity corresponding to the redshift is the cz value. The FWHM of the broad components of the $\mathrm{H\alpha}$ and $\mathrm{H\beta}$ recombination lines are $\mathrm{1715\pm14~km~s^{-1}}$, $\mathrm{5317\pm56~km~s^{-1}}$, $\mathrm{1944\pm55~km~s^{-1}}$, and $\mathrm{2507\pm31~km~s^{-1}}$, respectively. The FWHM of the narrow components of the $\mathrm{H\alpha}$ and $\mathrm{H\beta}$ recombination lines are $\mathrm{256\pm3~km~s^{-1}}$ and $\mathrm{335\pm6~km~s^{-1}}$, respectively. These FWHM values fall in the range of $\mathrm{1000-6000~km~s^{-1}}$ for broad HI lines in Seyfert 1 galaxies, and $\mathrm{300-800~km~s^{-1}}$ for narrow lines in both classes of Seyfert galaxies, as specified in \citet{Osterbrock}.
   
   \begin{table*}
   \caption{Values for Nucleus A}
   \label{table:flux_a}
   \centering
   \scalebox{0.8}{
   \begin{tabular}{c c c c c c c}
   \hline \hline 
   Emission Line & Observed Wavelength & Uncertainty in Wavelength & Flux Value & Uncertainty in Flux & FWHM & Uncertainty in FWHM \\
     & ($\AA$) & ($\AA$) & ($\mathrm{10^{-16}~erg~s^{-1}~cm^{-2}}$) & ($\mathrm{10^{-16}~erg~s^{-1}~cm^{-2}}$) & ($\mathrm{km~sec^{-1}}$) & ($\mathrm{km~sec^{-1}}$)\\
                \hline 
                $[\ion{O}{II}]\lambda3727$ & 4204.01 & 0.05 & 7.45 & 0.44 & 438 & 9\\
                $\mathrm{H\gamma_{\rm narrow}}$ & 4894.82 & 0.09 & 3.55 & 0.54 & 330 & 16\\ 
                $\mathrm{H\gamma_{\rm broad}}$ & 4903.44 & 0.52 & 2.01 & 0.43 & 3652 & 86\\
                $\mathrm{H\beta_{\rm narrow}}$ & 5482.47 & 0.04 & 8.99 & 0.49 & 335 & 6\\
                $\mathrm{H\beta_{\rm broad1}}$ & 5477.49 & 0.24 & 9.84 & 0.28 & 1944 & 55\\
                $\mathrm{H\beta_{\rm broad2}}$ & 5510.64 & 0.29 & 4.84 & 0.06 & 2507 & 31\\
                $[\ion{O}{III}]\lambda4959$ & 5591.68 & 0.06 & 6.27 & 0.55 & 407 & 8\\
                $[\ion{O}{III}]\lambda5007$ & 5645.88 & 0.02 & 20.1 & 0.23 & 430 & 3\\
                $[\ion{He}{I}]\lambda5876$ & 6630.27 & 1.09 & 4.35 & 0.04 & 3022 & 94\\
                $[\ion{O}{I}]\lambda6300$ & 7105.775 & 0.51 & 0.79 & 0.07 & 202 & 49\\
                $[\ion{N}{II}]\lambda6548$ & 7385.54 & 0.09 & 6.89 & 0.15 & 287 & 14\\
                $\mathrm{H\alpha_{\rm narrow}}$ & 7401.65 & 0.01 & 31.9 & 0.16 & 256 & 3\\
                $\mathrm{H\alpha_{\rm broad1}}$ & 7397.88 & 0.10 & 38.42 & 0.42 & 1715 & 14\\
                $\mathrm{H\alpha_{\rm broad2}}$ & 7410.05 & 0.45 & 41.88 & 0.86 & 5317 & 56\\
                $[\ion{N}{II}]\lambda6583$ & 7425.12 & 0.03 & 14.2 & 0.68 & 192 & 5\\ 
                $[\ion{S}{II}]\lambda6716$ & 7574.90 & 0.09 & 4.39 & 0.21 & 235 & 17\\
                $[\ion{S}{II}]\lambda6731$ & 7591.79 & 0.09 & 4.45 & 0.05 & 248 & 18\\
   \hline
   \end{tabular}}
   \tablefoot{All uncertainties are the $\mathrm{1~\sigma}$ standard error values. The values stated are calculated for an aperture of 1~arcsec.}
   \end{table*}
   
   \subsection{Optical spectrum of Nucleus B}
   
   \begin{figure*}[h]
     \centering
     \includegraphics[width=1\textwidth]{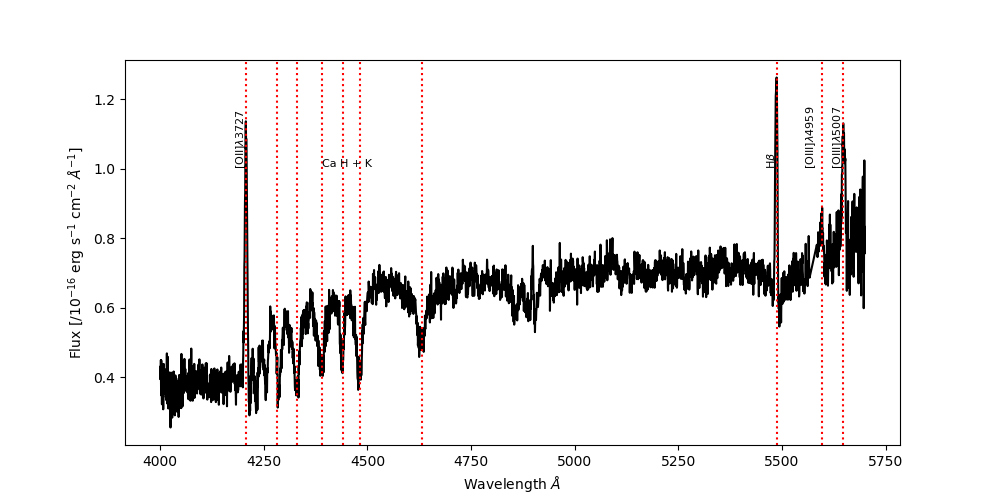}
     \caption{One-dimensional observed optical spectrum of SDSSJ134420.86+663717.8 nucleus B in the wavelength range $4000-5700~\AA$. The observed emission and absorption lines have been marked by vertical red dotted lines and named. The aperture size for spectral extraction is 1 arcsec. The wavelength axis depicts the observed wavelength.}%
     \label{fig:nuc_b_b}
   \end{figure*}
   
   \begin{figure*}[h]
     \centering
     \includegraphics[width=1\textwidth]{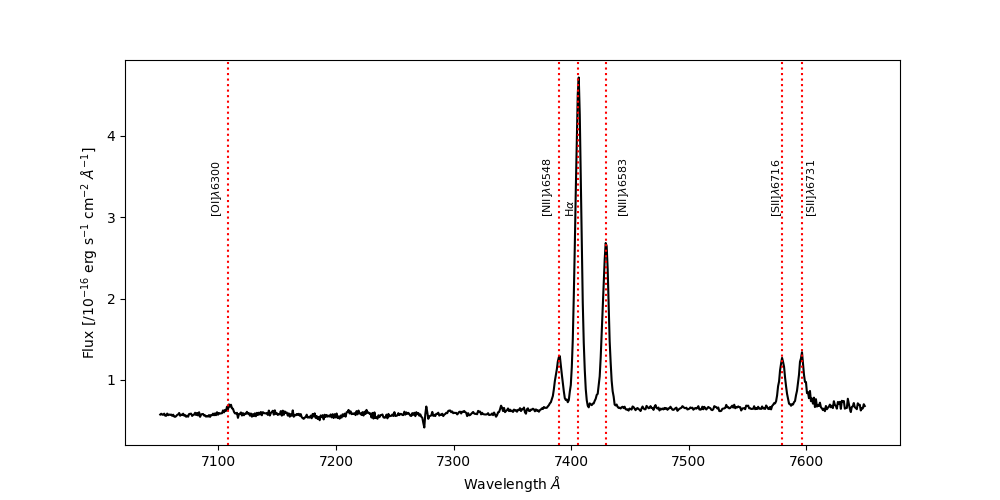}
     \caption{One-dimensional observed optical spectrum of SDSSJ134420.86+663717.8 nucleus B in the wavelength range $7000-7700~\AA$. The observed emission and absorption lines have been marked by vertical red dotted lines and named. The aperture size for spectral extraction is 1 arcsec. The wavelength axis depicts the observed wavelength.}%
     \label{fig:nuc_b_r}
   \end{figure*}

   Figures \ref{fig:nuc_b_b} and \ref{fig:nuc_b_r} show the optical spectrum of nucleus B. Several stellar absorption features are clearly visible along with some emission lines. Nucleus B shows fewer emission lines compared to nucleus A. No broad emission features are noticeable. Table \ref{table:flux_b} lists the different emission lines, their flux and FWHM values. The FWHM of the narrow $\mathrm{H\alpha}$ and $\mathrm{H\beta}$ recombination lines are $\mathrm{250\pm2~km~s^{-1}}$ and $\mathrm{256\pm16~km~s^{-1}}$, respectively. Some stellar absorption can be see at the $\mathrm{H\beta}$ line. The stellar absorption lines were fit using the 'penalised Pixel-Fitting' (pPXF, \citep{Cappellari, Cappellarib} method, and were taken into consideration while estimating emission line fluxes (see also, Section \ref{pPXF}). The average redshift calculated from the central positions of various emission lines is $\mathrm{0.1285\pm0.0001}$. The plots of spatial position versus intensity for $\mathrm{H\alpha}$ and the continuum show that both of them peak at the same point, which we consider to be the position of the central nucleus. We use this to estimate that the systemic redshift of Nucleus B is the redshift value displayed by the Hydrogen recombination lines at the central position of Nucleus B, that is, $\mathrm{0.1285}$. However, interestingly, the redshift of the $[\ion{O}{III}]\lambda5007$ is smaller at $\mathrm{0.1282}$ ($\sim38460~\mathrm{km~s^{-1}}$), whereas the redshift of the $[\ion{O}{II}]\lambda3727$ line is larger at $\mathrm{0.1288}$ ($\sim 38640~\mathrm{km~s^{-1}}$) compared to the mean redshift of 0.1285 ($\sim 38550~\mathrm{km~s^{-1}}$). The velocities corresponding to the redshifts are the cz values. $[\ion{O}{II}]\lambda3727$ is considered to be a good tracer for star formation. However, \citet{yan} state that [\ion{O}{II}] emission in galaxies could have other possible sources like AGN, fast shock waves, post-asymptotic giant branch (AGB) stars, and cooling flows. The comparatively higher value of redshift could arise from star formation in cool molecular clouds accumulated due to cooling flows. \citet{boller} categorise SDSSJ134420.86+663727.8 as a source of diffuse X-ray emission. The X-ray emission is suggested to be indicative of cooling flows. If the cooling flow is rapid, an inflow of gas is created to satisfy the condition of hydrostatic equilibrium. The lower redshift associated with the $[\ion{O}{III}]\lambda5007$ line could arise due to emission resulting from an outflow. On checking the value of its FWHM, we notice that it is $\mathrm{429\pm35~km~s^{-1}}$, which is higher than the FWHM values of the H$\alpha$ and H$\beta$ lines. The $[\ion{O}{III}]\lambda5007$ emission line is usually associated with outflows in the narrow line region (NLR). In such cases, the $[\ion{O}{III}]\lambda5007$ profile exhibits a blue asymmetry, in that the peak of the line is skewed blue-ward. However, in our case, the $[\ion{O}{III}]\lambda5007$ line is not reliable since it falls towards the end of the blue spectrum, and is very noisy and weak. Thus, we refrain from drawing any concrete conclusions based on the redshift and shape of the $[\ion{O}{III}]\lambda5007$ line.
   
   \begin{table*}
   \caption{Values for Nucleus B}             
   \label{table:flux_b}      
   \centering  
   \scalebox{0.8}{
   \begin{tabular}{c c c c c c c}        
   \hline\hline  
   Emission Line & Observed Wavelength & Uncertainty in Wavelength & Flux Value & Uncertainty in Flux & FWHM & Uncertainty in FWHM \\
     & ($\AA$) & ($\AA$) & ($\mathrm{10^{-16}~erg~s^{-1}~cm^{-2}}$) & ($\mathrm{10^{-16}~erg~s^{-1}~cm^{-2}}$) & ($\mathrm{km~sec^{-1}}$) & ($\mathrm{km~sec^{-1}}$) \\
            \hline $[\ion{O}{II}]\lambda3727$ & 4206.95 & 0.12 & 1.50 & 0.27 & 383 & 20\\
            $\mathrm{H\beta_{\rm narrow}}$ & 5486.34 & 0.13 & 1.35 & 0.16 & 256 & 16\\
            $[\ion{O}{III}]\lambda5007$ & 5648.63 & 0.28  & 0.62 & 0.06 & 429 & 35\\
            $[\ion{O}{I}]\lambda6300$ & 7108.99 & 0.64 & 0.19 & 0.02 & 404 & 65\\
            $[\ion{N}{II}]\lambda6548$ & 7389.56 & 0.11 & 1.14 & 0.04 & 323 & 10\\
            $\mathrm{H\alpha_{\rm narrow}}$ & 7406.12 & 0.01 & 7.51 & 0.32 & 250 & 2\\
            $[\ion{N}{II}]\lambda6583$ & 7429.38 & 0.03 & 3.78 & 0.12 & 265 & 3\\ 
            $[\ion{S}{II}]\lambda6716$ & 7579.51 & 0.11 & 1.14 & 0.02 & 241 & 9\\
            $[\ion{S}{II}]\lambda6731$ & 7596.16 & 0.10 & 1.09 & 0.07 & 273 & 10\\
            \hline                                   
   \end{tabular}}
   \tablefoot{All uncertainties are the $\mathrm{1~\sigma}$ standard error values. The values stated are calculated for an aperture of 1~arcsec.}
   \end{table*}
   
   \section{Discussion}\label{4}
   Once we obtained the optical spectra of both the galaxies along with the flux values of the different emission lines, we could use this information to deduce some properties associated with the individual galaxies. We discuss our findings in the subsections below. Having this knowledge at hand is useful to reconstruct the picture of the interaction (e.g., while using simulations). 
   \subsection{Diagnostic diagrams}
\begin{figure}[h!]
  \centering
  \includegraphics[width=0.50\textwidth]{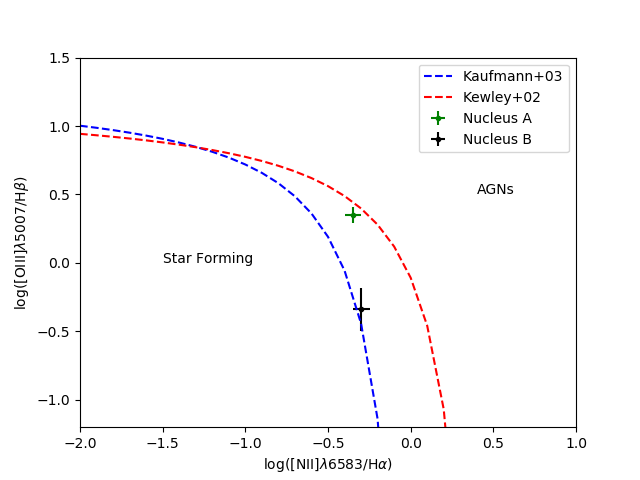}
  \centering
  \includegraphics[width=0.50\textwidth]{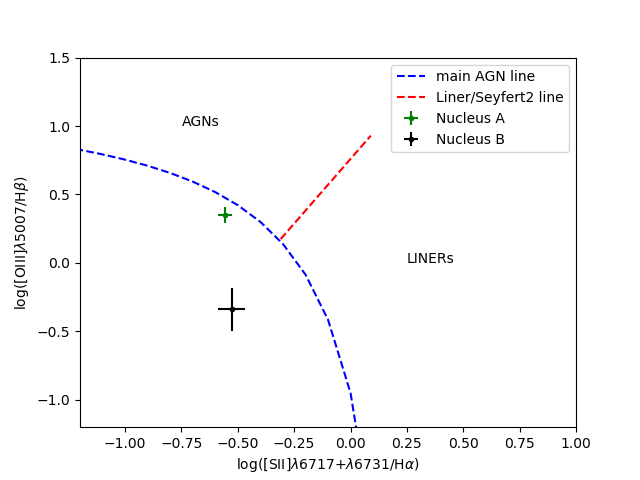}
  \centering
  \includegraphics[width=0.50\textwidth]{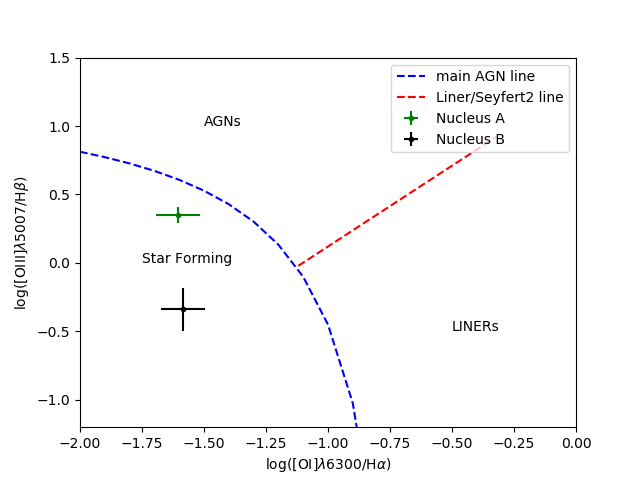}
  \caption{\label{fig:bpt_1} Plots of $\log([\ion{O}{III}]\lambda5007/\mathrm{H\beta}$) vs $\log([\ion{N}{II}]\lambda6583/\mathrm{H\alpha}$), $\log([\ion{O}{III}]\lambda5007/\mathrm{H\beta}$) vs $\log([\ion{S}{II}](\lambda6717+\lambda6731)/\mathrm{H\alpha}$) and $\log([\ion{O}{III}]\lambda5007/\mathrm{H\beta}$) vs $\log([\ion{O}{I}]\lambda6300/\mathrm{H\alpha}$). The green point represents nucleus A and the black point represents nucleus B. The data points have been calculated for an aperture of 1~arcsec.}
  \end{figure}

    We use the diagnostic diagram originally proposed by Baldwin, Phillips and Terlevich \citep{Baldwin}, along with two other diagnostic diagrams which are discussed by \citep{Veilleux1987}. The diagnostic diagrams are essentially a set of three different plots of the ratios of the fluxes of emission lines that are close enough to each other in wavelength to render reddening ineffective. They are used to determine the excitation mechanisms of the gas, and thereby they help us to distinguish between AGN and star-forming galaxies \citep{Kewley2019}. The different excitation mechanisms responsible for line emission could be normal $\ion{H}{II}$ regions, planetary nebulae, objects photoionised by a harder radiation field, AGN, shocks, and low-ionisation nuclear emission-line region (LINER) emission. The Equations~(\ref{eqn1}, \ref{eqn2}, and \ref{eqn3}) \citep{Groves} define the curves in the diagrams which separate the AGN from star-forming galaxies (see the plots in Figure \ref{fig:bpt_1}). The emission lines used are $[\ion{O}{III}]\lambda5007$, $\mathrm{H\beta}$, $[\ion{N}{II}]\lambda6583$, $\mathrm{H\alpha}$, $[\ion{O}{I}]\lambda6300$, $[\ion{S}{II}]\lambda6716$ and $[\ion{S}{II}]\lambda6731$. 
   \begin{equation}\label{eqn1}
   \log([\ion{O}{III}]/\mathrm{H\beta}) = 0.61/{(\log([\ion{N}{II}]/\mathrm{H\alpha}) - 0.47)} + 1.19. \\
   \end{equation}
   \begin{equation}\label{eqn2}
   \log([\ion{O}{III}]/\mathrm{H\beta}) = 0.72/{(\log([\ion{S}{II}]/\mathrm{H\alpha}) - 0.32)} + 1.30). \\
   \end{equation}
   \begin{equation}\label{eqn3}
   \log([\ion{O}{III}]/\mathrm{H\beta}) = 0.73/{(\log([\ion{O}{I}]/\mathrm{H\alpha}) + 0.59)} + 1.33. \\ 
   \end{equation}
   Additionally, there is a curve~(Equation \ref{eqn4}) proposed by \citet{Kauffmann}, which provides another delineation between AGN and star-forming galaxies. The curve (Equation \ref{eqn1}) proposed by \citet{Kewley} is called the maximum starburst line and is based on theoretical modelling, while the curve (Equation \ref{eqn4}) put forth by \citet{Kauffmann} is called the pure star-forming line and is based on observational data. If an object falls in the space between these two curves, they are considered to be composite or transitioning objects. 
   \begin{equation}\label{eqn4}
   \log([\ion{O}{III}]/\mathrm{H\beta}) = 0.61/(\log([\ion{N}{II}]/\mathrm{H\alpha}) - 0.05)) + 1.3. \\
   \end{equation}
   To further classify the AGN as Seyferts or LINERS, there are two more curves~(Equations \ref{eqn5} and \ref{eqn6}). The LINER region can also be populated by regions of shocked gas \citep{Rich2014}.
   \begin{equation}\label{eqn5}
   \log([\ion{O}{III}]/\mathrm{H\beta}) = 1.89\log([\ion{S}{II}]/\mathrm{H\alpha}) + 0.76. \\
   \end{equation}
   \begin{equation}\label{eqn6}
   \log([\ion{O}{III}]/\mathrm{H\beta}) = 1.18\log([\ion{O}{I}]/\mathrm{H\alpha}) + 1.30. \\
   \end{equation}
   
  Figure \ref{fig:bpt_1} shows the three different plots. We can see from the first plot in Figure \ref{fig:bpt_1} that while both nuclei lie in the composite/transition phase, nucleus A is very close to the AGN part, and nucleus B lies very close to the star forming part of the composite or transition phase. This could be due to the interaction of the galaxies. Interactions of galaxies cause huge amounts of gas and dust to get funneled through to the centers of the galaxies causing star-formation and might make the super-massive black holes (SMBH) at the centres of galaxies active, in the sense that they start accreting gas and dust. The position of the two galaxies in the transitioning phase is further bolstered by the remaining two figures (see the second and third plot in Figure \ref{fig:bpt_1}), where nucleus A lies in the star-forming region close to the dividing curve, while nucleus B lies much further down, with low ionisation values. The [$\ion{O}{III}]\lambda5007$ lying at the red-ward end of the MODS Blue spectrum for nucleus B is very noisy, this is reflected in the uncertainty in the positions of the data points. Similarly, the [$\ion{O}{I}$]$\lambda6300$ line has very low flux values and has a higher uncertainty in its measurement. Thus, we can conclude that while nucleus A has definite AGN like properties, nucleus B is quiescent, with low ionisation levels, and could be star-forming. This is in agreement with the conclusions we drew in Section \ref{section3} from looking at the emission lines displayed by both the nuclei.  
  
  \subsection{Gas rotation curve}
  \begin{figure*}[h]
  \centering
  \includegraphics[width=1\textwidth]{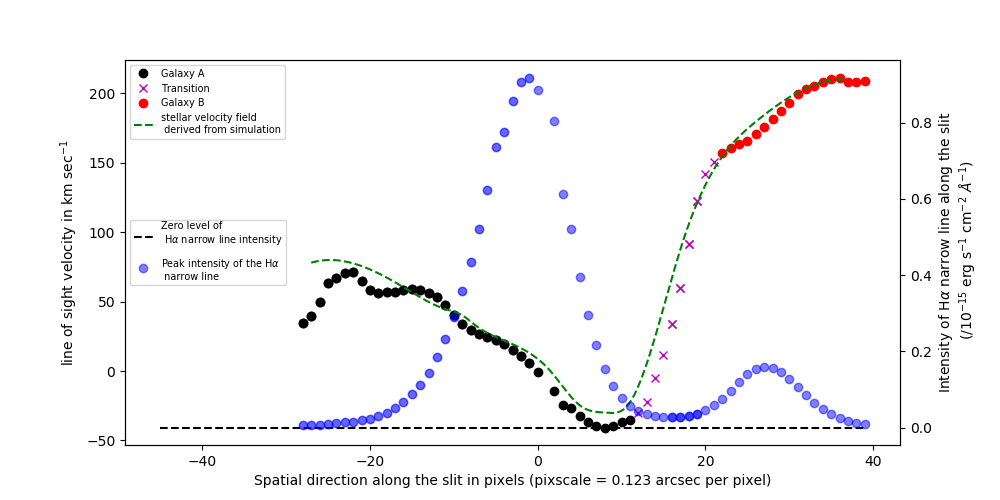}
  \caption{\label{fig:rot_curve}Line-of-sight velocity in $\mathrm{km~s^{-1}}$ calculated from the H$\alpha$ narrow line vs the radial distance from the centre of nucleus A in pixels. The pixel scale is 0.123 arcsec per pixel. Overplotted is the intensity profile of the $\mathrm{H\alpha}$ narrow line. The black dashed line represents the $\mathrm{intensity = 0}$ level for the $\mathrm{H\alpha}$ line. The zero of the x-axis represents the point of highest intensity of the $\mathrm{H\alpha}$ narrow line. The dashed green line represents the stellar velocity field derived from our simulations explained in Section \ref{vel_field}.}
  \end{figure*}
  
  The rotation curve of a galaxy is the plot of line-of-sight (LOS) velocity against the radial distance from the centre of the same \citep{Jog}. The rotation curve helps in understanding the distribution of mass as a function of the distance from the centre of the galaxy.
  
  Figure \ref{fig:rot_curve} shows the gas rotation curves for the nuclei A and B along the aperture of approximately 8~arcsec on the slit. Additionally, we mark the velocities obtained from our simulation with dashed green lines, to aid in comparing the results of our simulation with the observational data. See Section \ref{vel_field} for more details about the velocity field of the simulated model. It can be seen from the SDSS image of SDSS~J134420.86+663717.8 (see Figure \ref{fig:sdss}) that the galaxies appear to be significantly overlapping. As a result, the extended gas disks associated with the galaxies might overlap in projection. Correspondingly, there is a slight broadening seen in some narrow emission lines, however, no split in profiles is noticed. The rotation curve seems to be divided into three distinct parts. The black dots represent the velocities associated with nucleus A, the red dots represent the velocities associated with nucleus B, while the magenta crosses represent the transition phase between nucleus A and nucleus B. The transition phase is defined as the region between the nuclei along the slit, where the LOS velocity associated with the $\mathrm{H\alpha}$ narrow line increases linearly with spatial distance. The velocities are calculated from the shift of the centre of the $\mathrm{H\alpha}$ narrow line along the slit using the Gaussian fits explained in Section~\ref{section3}. In blue, we have overplotted the intensity profile of the H$\alpha$ narrow line along the slit to provide a better understanding about the nature of the rotation curve. The black dashed line represents the level at which there would be no flux from the $\mathrm{H\alpha}$ line. The zero on the x-axis was arbitrarily selected to coincide with the point on the slit with the highest intensity of the $\mathrm{H\alpha}$ narrow line. The shape of the rotation curve suggests that the galaxies are rotating in opposite directions. This counter-rotation is a feature that can be seen in many cases of interacting galaxies. It arises due to the initial spins of the progenitor galaxies \citep{Barnes2016}. The range of velocities for nucleus A is approximately $\mathrm{\pm50~km~s^{-1}}$, while for nucleus B, the range of velocities is approximately $\mathrm{\pm25~km~s^{-1}}$. The shape of the rotation curve shows some asymmetry in the case of galaxy A (since we are no longer dealing with just the central arcsec but rather a longer aperture, we refer to the galaxy associated with nucleus A as galaxy A and the galaxy associated with nucleus B as galaxy B). Asymmetry due to interaction with neighbouring galaxies is a well documented phenomenon \citep{Hodge, Vulcani}. Galaxy B has a uniform rotation curve, with no extraordinary features. The transition phase appears to be different from the two other sections, with the LOS velocity increasing almost linearly with spatial distance along the slit. The spatial width of the transition region is approximately equal to the point source function (PSF) of the telescope, thus, we conclude that the transition region has contribution from both of the nuclei. We note that while the flux from the $\mathrm{H\alpha}$ line is quite low in the region between the two nuclei, it is never zero. The $\mathrm{H\alpha}$ emission line is always present in the transition phase and it is possible to perform Gaussian fitting.
  
  The velocities obtained from the rotation curve can be used to estimate the masses of the central bulges of the galaxies. For this purpose, we used the Baryonic Tully-Fisher relation (BTFR). The original Tully-Fisher relation \citep{Tully} is the relation between the luminosity and the rotation velocity of a galaxy, wherein luminosity is considered to be a proxy for total mass of the system. However, luminosity varies with galaxy type, leading to slightly different relations depending on the galaxy being considered. \citet{McGaugh, Torres-Flores, McGaugha} (and references therein) show that considering baryonic mass, that is, gas mass as well as stellar mass, leads to a more accurate relation between galactic mass and the rotation velocity. We use the maximum velocity obtained from the rotation curve as seen in Figure \ref{fig:rot_curve} to calculate the masses of the bulges in the central $\mathrm{2\arcsec}$. For this purpose, we used the relation stated in \citet{Torres-Flores}:
  \begin{equation}
      {M_\mathrm{bar} = 10^{(2.21\pm0.61)}\mathrm{V_{max}}^{(3.64\pm0.28)}}.
  \end{equation}
  The relation yields a baryonic bulge mass of approximately $\mathrm{2\times10^8~M_\odot}$ for galaxy A and a baryonic bulge mass of approximately $\mathrm{1\times10^7~M_\odot}$ for galaxy B. However, we note that the calculated values of the baryonic bulge masses are an underestimate as we cannot be sure if the slit direction probes the major axis of rotation, and considering that the galaxies are interacting, there is a likelihood of perturbation in the velocities, such that the gradient of the velocity lowers. Additionally, nucleus B could have a higher inclination than nucleus A, making it edge-on to our view as compared to nucleus A, which seems to be face-on based on its spectrum. Taking this into consideration, if we divide the value of the baryonic mass of the bulge of galaxy B with the sine value of its inclination, we get approximately equal bulge mass values of $\mathrm{10^8~M_\odot}$ for both of the galaxies. The value of the inclination is a rough estimate based on the knowledge we have from the observational results and the results of the simulation (see Section~\ref{4.5} for a detailed description of the simulation). The inclinations were read off the three dimensional model cube and not taken from the list of simulation parameters. We thus conclude that nucleus A and nucleus B have almost equally massive bulges. 
  
  \subsection{Mass of the central supermassive black holes}
  
  The nuclear spectrum can also be used to calculate the mass of the central super-massive black hole of the galaxy. The mass can be calculated in various ways, depending upon the spectral features available, for example, the virial method (explained in detail in the next subsection), the $M_\mathrm{{BH}}-\sigma_*$ relation, the $M_\mathrm{{BH}}-M_\mathrm{{bulge}}$ relation, and the $M_\mathrm{{BH}}-L_\mathrm{{bulge}}$ relation to name a few (\cite{Busch} and references therein). In this case, we used the virial method to calculate the mass for nucleus A since we detect broad emission lines, and the $M_\mathrm{{BH}}-\sigma_*$ relation to calculate the mass for nucleus B, since the spectrum for nucleus B shows stellar absorption lines. 
  
  \subsubsection{Mass of the black hole for nucleus A}
  The broad lines in the optical spectrum of nucleus A can be used to calculate the mass of the SMBH at its center. The broad lines arise from the broad line region (BLR) around the central SMBH, where gas rotates at high velocities. Consequently, we can use the virial theorem to derive the mass of the central object.
  
  For the case of AGN with broad lines, Equation~\ref{eqn9} can be used to calculate the mass of the central black hole in the following form \citep{Peterson},
  \begin{equation}\label{eqn9}
   M_\mathrm{BH} = f \frac{\mathrm{\Delta V^{2}}R_\mathrm{BLR}}{G},  \\
  \end{equation}
  
  where $\mathrm{\Delta V}$ is the velocity of the broad line region gas, usually characterised by the FWHM or the line dispersion ($\sigma$) of the $\mathrm{H\beta}$ broad line, $R_\mathrm{{BLR}}$ is the size of the BLR and $G$ is the gravitational constant. We used the FWHM values obtained from the Guassian fit procedure to calculate the mass of the central SMBH. The size of the BLR and the luminosity of the AGN are related to each other by an approximate relation of the form: $R_\mathrm{{BLR}}\propto L^{1/2}$ \citep{Shields}. Thus, the luminosity of the AGN continuum can be used to represent the extent of the BLR. The dimensionless factor, '$f$', called the virial factor, subsumes within it the effects of everything concerning BLR geometry, its inclination, the kinematics that govern the region, and everything else that is not known to us. The value of '$f$' is different for every individual AGN but is expected to be of the order of unity. An average value of '$f$' can be calculated by normalising the AGN $M_{BH}-\sigma_*$ relationship to that of quiescent galaxies \citep{Peterson}. 
  
  Since the $\mathrm{H\beta}$ line in our data falls towards the end of the blue spectrum, where the dichroic splits the light into blue and red components, it is better to use the line width and the luminosity of the $\mathrm{H\alpha}$ line instead. The relations used to relate the luminosity and the FWHM of the H$\alpha$ line to the continuum luminosity and the FWHM of the $\mathrm{H\beta}$ line are taken from \citet{Greene} and \citet{Woo}. Finally, the equation used to calculate the mass of the black hole is:
  \begin{equation}\label{eqn13}
  M_\mathrm{{BH}} = f\times10^{6.544} \Bigg(\frac{L_{\mathrm{ H\alpha}}}{10^{42}~ \mathrm{erg~s^{-1}}}\Bigg)^{0.46} \Bigg(\frac{FWHM_{\mathrm{ H\alpha}}}{10^3~ \mathrm{km~s^{-1}}}\Bigg)^{2.06}~M_\odot. \\
  \end{equation}
  The value of $f$ for FWHM-based method was adopted to be 1.12 \citep{Woo}. The FWHM of the broad H$\alpha$ line was taken to be the FWHM of the overall broad component from the H$\alpha$ + $[\ion{N}{II}]$ complex, while the luminosity was calculated as the sum of the luminosities of the two broad components. With the Hubble constant taken to be $\mathrm{H_0=69.6~km~s^{-1}}$, the mass of the central SMBH is calculated to be approximately $2\times10^7~M_\odot$. We refrain from stating formal error estimates since the relations we used have intrinsic scatter and the value should be taken to be an order of magnitude estimate.
  
  \subsubsection{Mass of the black hole for nucleus B}\label{pPXF}
  In the absence of broad line features, the $M_\mathrm{{BH}}-\sigma_*$ relation can be used to estimate the mass of the central black hole of a galaxy, where $\sigma_*$ is the stellar velocity dispersion of the galactic bulge. The $M_\mathrm{{BH}}-\sigma_*$ relation, called the Faber Jackson law for black holes \citep{Merritt}, has undergone several changes over the years with regards to the power of $\sigma_*$. The relation used here was put forth by \citet{McConell} and takes the form:
  \begin{equation}\label{eqn17}
  \frac{M_\mathrm{{BH}}}{10^8 M_\odot} \approx 1.9 \Bigg(\frac{\sigma_*}{200~ \mathrm{km~s^{-1}}}\Bigg)^{5.1}.  \\
  \end{equation}
  The stellar velocity dispersion of the galaxy was found by fitting a stellar template to the optical spectrum of the galaxy using the pPXF package (version v6.7.6), developed by Michele Cappellari \citep{Cappellari, Cappellarib}. The pPXF package can fit a large set of stellar templates without encountering a template mismatch due to different ranges of wavelength. We used the stellar spectra from the MILES library \citep{Vazdekis} to fit the stellar template to our data. The MILES library consists of synthetic single-age, single-metallicity stellar populations over the full optical spectral range. The value of $\sigma_*$ obtained from the fitting was $139\pm32~km~s^{-1}$. Plugging this in Equation \ref{eqn17}, the mass of the black hole for nucleus B is calculated to be approximately $3\times10^7~M_\odot$. As with nucleus A, the value for the mass of the central SMBH should be taken as an order of magnitude estimate since the relations we use have intrinsic scatter which deters us from making a formal error estimation.
  
  \subsection{Masses of the galaxies}\label{section4.4}
  Finally, we used mass to light ratio to estimate the masses of the individual galaxies. To do this, we used an average mass to light ratio of $\sim~200~h~ M_\odot/L_\odot$ from \citet{Gonzalez2019}, for close galaxy pairs in the optical regime. We used the continuum luminosities calculated at various distances from the centre to fit Sersic profiles for both of the galaxies, and used these to get integrated values of the luminosities of the entire galaxies. Then we plugged this value of luminosity in the mass to light ratio to estimate the mass of the galaxy. We excluded the central arcsec while fitting the Sersic profile in order to disconsider the contribution from the centre of the galaxy. This yielded values of approximately $2\times10^{12}~M_\odot$ for galaxy A and approximately $6\times10^{11}~M_\odot$ for galaxy B. These values seem to be in agreement with the general galaxy mass estimates. However, we note that as the galaxies are interacting with each other, there must be intermixing of gas, which is not properly accounted for here. This can lead to an overestimation or underestimation of mass as we cannot be sure whether the contribution to luminosity at any particular point comes from one or two galaxies. Consequently, the values of mass quoted here should be taken as order of magnitude estimates. 
  
  \section{Simulation}\label{4.5}
  While we could ascertain some of the properties of the galaxies from the data available to us, it is interesting to understand the initial conditions of the participant galaxies. For this, simulations play a big role in increasing our understanding. The current morphology of the system provides an inkling regarding the angles at which the galaxies might have approached each other. We must keep in mind, though, that since what we see is the two-dimensional projection of the three-dimensional morphology on the plane of the sky, one projection might be explained by many different 3D models. Projection effects could cause tidal tails to cross and might lead to association with the 'incorrect' galaxy, for example, Antennae galaxies have crossing tidal tails. Additionally, one model might be used to explain different structures by introducing different viewing angles, for example, \citet{Schawachter2004} report a multi-particle model for the QSO 3C48 with conditions similar to those used to interpret the appearance of the Antennae galaxies but with a different viewing angle. Our aim in building the simulation is to model a spatial and kinematical look-alike of SDSS~J134420.86+663717.8, and shed some light on one of the possible sets of initial conditions.
  
  \subsection{Basis of our simulation}
  A visual inspection of the image of the object while keeping in mind the properties determined from the analysis in Section~\ref{4} supplied a vague idea regarding the initial conditions. The galaxies are at different redshifts and have almost equally massive central SMBHs. The tidal tail of the galaxy at higher redshift appears elongated and extended, while the tidal tail of the galaxy at lower redshift appears to be foreshortened in comparison. The foreshortening might be a result of the viewing angle being such that we look at the tidal tail edge-on. With this in mind, we looked at a model of interacting galaxies presented by \citet{Barnes} and \citet{Barnesb}. They presented eight different sets of encounter parameters and studied the outcomes for the case of two equal mass galaxies.
  
  The models consisted of three types of matter, namely stars, dark matter, and interstellar gas. \citet{Barnes} assumed that the stars and the dark matter could be described by the collisionless Boltzmann equation, whereas the gas was modelled by the standard laws for a compressible fluid along with gravitational forces, radiative cooling, and shock heating. Poisson's equation was used to obtain the gravitational field generated by the stars, dark matter, and gas. The bulge to disk to halo mass ratio was taken to be 1:3:16, with the gas contributing 10\% of the disk mass. We focused on one of the eight scenarios presented in \citet{Barnes}: encounter A, as, coincidentally, the outcome of this scenario resembles SDSS J134420.86+663717.8 most closely. 
  
\begin{figure}[h]
  \centering
  \includegraphics[width=0.5\textwidth]{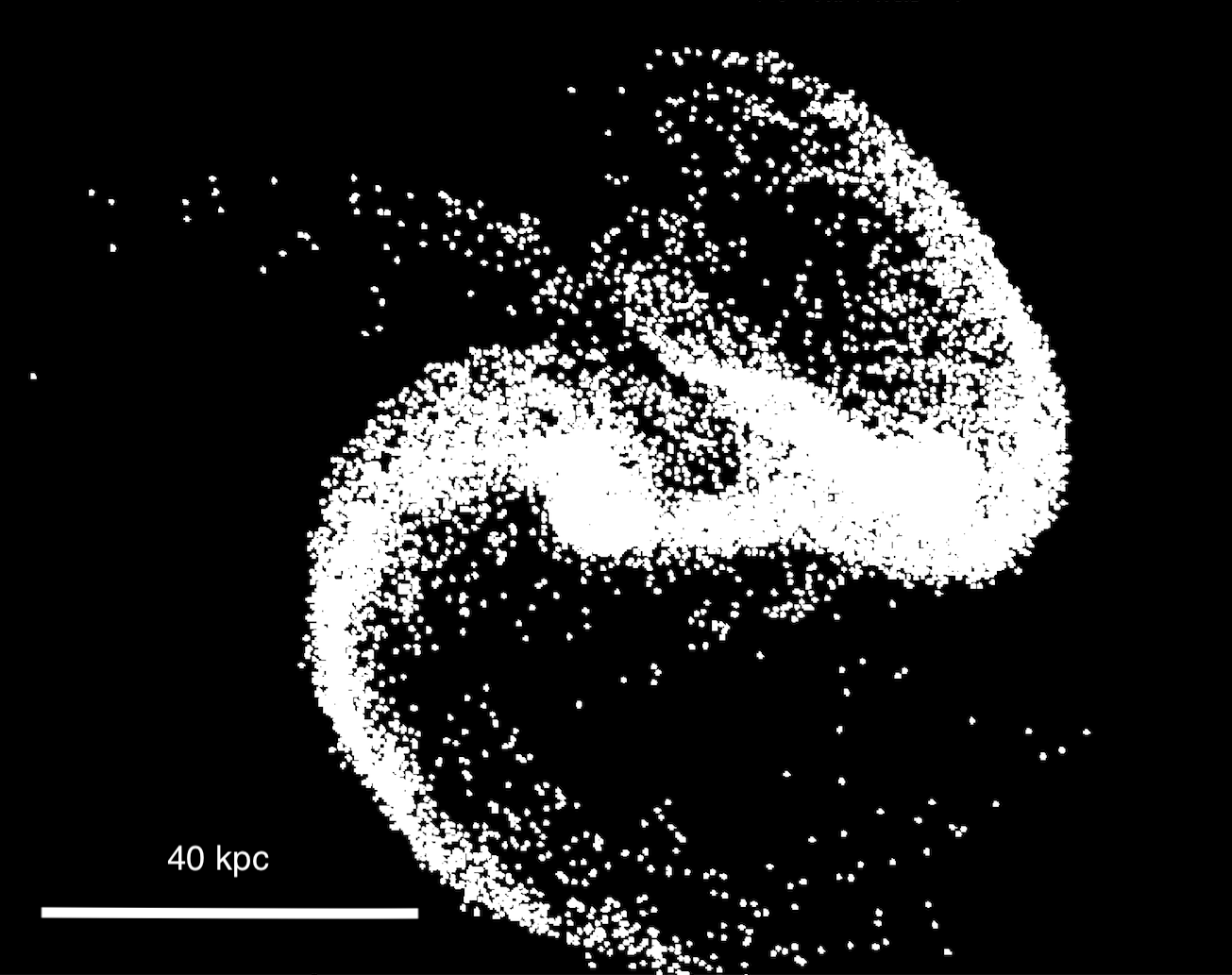}
  \caption{\label{fig:simulation}Simulated SDSSJ134420.86+663717.8 look-alike $\approx$ 375 Myr after the first collision. Comparison with Figure \ref{fig:sdss} shows striking similarities between SDSSJ134420.86+663717.8 and the model. Two distinct nuclei are visible, along with the extensive tidal tail associated with nucleus B, which is the nucleus to the left in the image above. Credit: \citep{Barnes}}
  \end{figure}

  The parameters for this parabolic encounter are: pericentric separation $R_\mathrm{p}=0.2$, $i_1=71$, $\omega_1=30$, $i_2=0$, $\omega_2=0$, and total number of particles $N=10,112$. The parameters, '$i$' and '$\omega$', describe the inclination of the orbit to the spin plane of the galaxy, and the angle between the node and the point of closest approach to the center of the disk without being disrupted, respectively, as defined and used by \citet{Toomreb}. They used a length unit of 40~kpc and a time unit of 250~Myr. The paper \citep{Barnes} shows pictorial representations of the objects at various time steps and explains each step in careful detail. Figure \ref{fig:simulation} shows their simulated result that resembles SDSS~J134420.86+663717.8 most closely. A comparison with Figure \ref{fig:sdss} reveals striking similarities in their morphology. However, there was still some difference in the morphology of the Barnes and Hernquist Model (hereafter referred to as the B\&H Model) and the SDSS image, for example, the line connecting the two nuclei in the B\&H Model should have a slight tilt to resemble the SDSS image, the tip of the tidal tail associated with nucleus B is on the same level as nucleus B in the B\&H Model, while this is not the case in the SDSS image, the tidal tail associated with nucleus A is much longer in the B\&H Model as compared to the SDSS image. Moreover, both of the tidal tails are equally prominent. This is not the case for SDSS~J134420.86+663717.8. Having been thus convinced about the similarities and the differences between the B\&H Model and the SDSS image, we used these parameters as a starting point in our N-body simulation to see if it would yield a better result with some modifications and help shed some light on the initial conditions of the two galaxies. We started with the parameters used by \citet{Barnes} and kept modifying them until we reached a point where the result of the simulation looked most alike SDSS ~J134420.86+663717.8. We present our results in the next section.
  
  \subsection{N-body simulation}
  In their paper, \citet{Toomreb} used three body calculations between the two central masses and each of the particles in the disk to demonstrate that bridges and tails can form due to tidal kinematic interactions. We used their model as a base, but in order to obtain more detailed knowledge of the orbital elements and the orientation of the hosts, we employed N-body calculations. A predictor-corrector method was used and implemented based on the description of the code and the algorithms in \citet{Aarseth}. 
  
  \subsubsection{Method}
  If we consider particles to be subject to gravitational forces only, the forces would vary smoothly with respect to time. In such a case, a polynomial fit could be applied to the force (or acceleration), and force at time $\mathrm{t_0+\Delta t}$ could be predicted via extrapolation. N-body simulations have been used to study a wide range of problems, for example: tidal disruptions of dwarf spheroid galaxies orbiting the Milky Way \citep{Oh}.
  
  Instead of initialising each point individually, we used twelve parameters to set the starting conditions of the system: accuracy parameter $\eta$, softening parameter or Plummer constant $\epsilon$, number of galaxies, number of particles $N$, outer radius, central mass, galactic mass ratios, velocity factors, rotations, translations, velocity transformations, and the output time step. The initial structure of our galaxies was that of two disc galaxies, with an outer disc which has the same mass as the central bulge. We set the number of particles $N$, number of rings, a radius of the outer ring, and a mass of the entire galaxy to generate our galaxy-like structure. For this case, we used an outer radius of 30~kpc, with twenty particles in the outer ring and ten particles in the inner ring, and an additional particle as the galactic centre, that is, bulge, molecular zone, and inner portion of the disc. The outer particles were spaced equally in an outer ring of radius $R=30~\mathrm{kpc}$, while the ten inner particles were spaced equally in an inner ring at radius $15~\mathrm{kpc}$, such that all particles had the same arc length separation to their respective 'left' and 'right' neighbours. The mass of the central particle was taken to be $10^{11}~M_\odot$ and is equal to the mass of all other particles. We do not explicitly consider a description of an extended mass distribution at the centre, like a bulge or a central molecular zone, due to the small number of particles, but approximate the mass distribution of the cloud through our central particle. This assumption, therefore, does not contradict our bulge mass estimates of approximately $\mathrm{10^{8}~M_\odot}$ for both of the galaxies, since the bulge mass values were derived for the central $2\arcsec$, while the central particle here spans approximately $5\arcsec$ in diameter. The values of the galaxy masses are in good agreement with the values obtained dynamically in Section~\ref{section4.4}. We initialised the galaxies with velocities of $\mathrm{50~pc~Myr^{-1}}$ for the blue (galaxy A) galaxy and $\mathrm{-125~pc~ Myr^{-1}}$ for the red (galaxy B) galaxy. We chose these values based on the velocity difference calculated from the rotation curve. However, we also tried larger and smaller values for the velocities of both galaxies in various combinations and concluded that the velocity values obtained from the observed data yield the most 'lookalike' results, spatially. Table \ref{table:Starting_Conditions} lists the starting parameters used to implement the simulation.
  
  \begin{table}
  \caption{Initialisation conditions for the simulation}
  \smallskip
  \centering
  \scalebox{0.8}{
  \begin{tabular}{c c c}
  \hline \hline Variable & Value & Units\\[0.7ex]
  \hline Number of galaxies & 2  & n.a\\
  N & [20,10] & n.a\\
  Outer radius & 30 & kpc\\
  Central mass & $\mathrm{10^{11}}$ & $\mathrm{M_\odot}$\\
  Galaxy mass ratios & [1, 1] & n.a\\
  \hline Rotations around x-axis & [$\mathrm{-17\pi/36}$,~$\mathrm{\pi/18}$] & radians\\
  Rotations around y-axis & [0,~0] & radians\\
  Rotations around z-axis & [$\mathrm{-\pi/18}$,~0] & radians\\
  \hline Galaxy 1 velocity factors & [1.32, 1.03] & n.a\\
  Galaxy 2 velocity factors & [1.32, 1.03] & n.a\\
  \hline Galaxy 1 translation & [-30, -25, 0] & kpc\\
  Galaxy 2 translation & [+30, +25, 0] & kpc\\
  \hline Galaxy 1 group velocity & [50, 5, 1] & pc/Myr\\ 
  Galaxy 2 group velocity & [-125, -5, -1] & pc/Myr\\
  \hline
  \end{tabular}}
  \label{table:Starting_Conditions}
  \end{table}
   
  \subsubsection{Limitations}
  The major limitation of our code is the rather small number of particles used to define our galaxies. However, we note that the total number of particles considered by us ($3~<~N_\mathrm{{total}}~<~100$) is within the range of 'typical particle number for applications' specified in \citet{Aarseth} (pg.~107). We will now present two different factors to demonstrate the accuracy of our results, that is, conservation of energy and stability of a galaxy without tidal forces. 
  
  Let us first consider the latter point: to be able to state with any certainty that the structure of the final simulation is due to galactic tidal interactions, we should check how the galaxy behaves in the absence of tidal forces. We define stable galaxies with three different characteristic times: $t_\mathrm{{start}}$: time at which we initialise the system, $t_\mathrm{{symmetry}}$: time at which the galaxy first loses its initial structure, and $t_\mathrm{{chaos}}$: time at which the galaxy first loses its symmetry. After $t_\mathrm{{chaos}}$ we cannot be certain that the galactic superstructures are caused by only tidal forces, they could as well be caused by the chaotic motion of the galaxies themselves. We refrain from crossing $t_\mathrm{{chaos}}$ ($t_\mathrm{{chaos}}=2500~\mathrm{Myr}$ in our case) in our analysis to guarantee the integrity of our results. Beyond $t_\mathrm{{chaos}}$, the system loses symmetry.
  
\begin{figure}[h]
  \centering
  \includegraphics[width=0.5\textwidth]{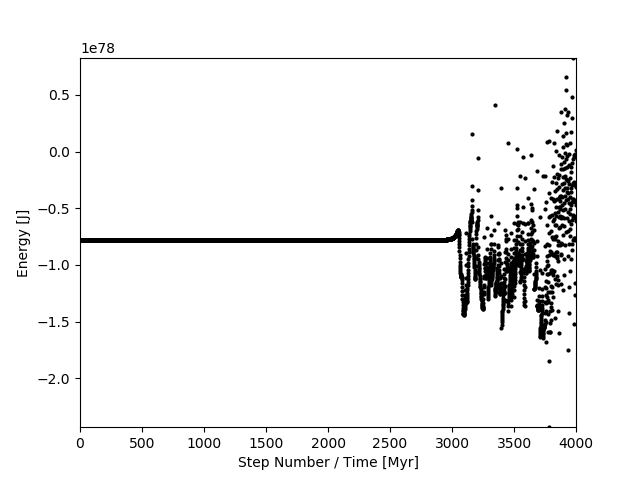}
  \caption{\label{fig:Energy}Time evolution of the energy of a single stable galaxy. There is an absolute break down of accuracy beyond $\mathrm{3000~Myr}$. Our simulation does not extend beyond $\mathrm{375~Myr}$ in time, which lies well within the time where energy conservation is rather accurate.}
  \end{figure}

  Going back to the former point concerning the conservation of energy, we consider the total energy of the system for the duration of the simulation. Figure \ref{fig:Energy} shows the time evolution of the energy of a single stable galaxy. There is a significant loss in accuracy after 3000~Myr. We do not consider any values beyond this time for our merger simulation, as we consider this point to be an absolute break-down for the accuracy of our simulation. For $t<3000~\mathrm{Myr}$, the maximum energy change is $\mathrm{1.81\times10^{-6}\%}$. We can compare this to the maximum energy change before $t_\mathrm{{chaos}}$ which is $\mathrm{1.22\times10^{-11}\%}$. In our analysis we never surpass $t_\mathrm{{chaos}}$. 
  
  \subsection{Visualisation}

Here, we show how the imaging and spectroscopy we present for this galaxy merger is matched by the simulations we performed.

    \subsubsection{Visualising the results from the simulation}
 In order to get a clear view of the distribution of matter predicted by the simulations and to have a higher degree of confidence regarding the structures involved, we run the simulation 15 times (each with 62 particles with 31 particles per galaxy) with minimal alterations to the starting conditions and overlap the resulting data to produce a single image with a total of 930 plotted points such that none of the single particles is of any particular interest. 
  
  \begin{figure}[h]
  \centering
  \includegraphics[width=0.38\textwidth]{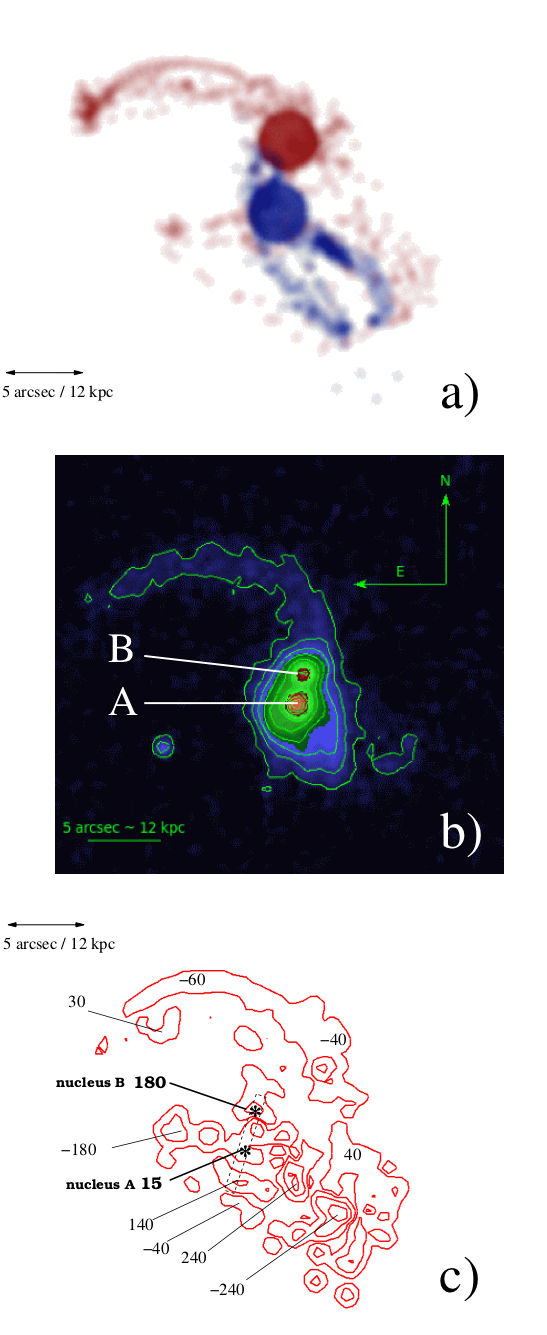}
  \caption{\label{fig:2d} Comparison of the SDSS image of the galaxies with the result of the simulation. The top panel (a) shows the model obtained using our N-body
simulation with the general region of nucleus A (blue) and B (blue) highlighted.
The middle panel (b) shows the optical SDSS image of galaxies
SDSS J134420.86+663717.8, superimposed with contours
starting at 10\% in steps of 10\%. Comparison of the
two figures shows striking similarities in their morphology,
especially in the shapes of the two tidal arms.
In the bottom panel (c), at
an angular resolution of about 1 arcsec, we show a smoothed velocity map obtained from the simulated data cubes. Contour levels are
-200, -100, -20, 20, 100 , 200  $\mathrm{km~s^{-1}}$. The dashed rectangular box shows the aperture along which the velocities used to plot the dashed green lines in the rotation curve in Figure \ref{fig:rot_curve} were extracted.}
  \end{figure}

  \begin{figure*}[h]
  \centering
  \includegraphics[width=0.75\textwidth]{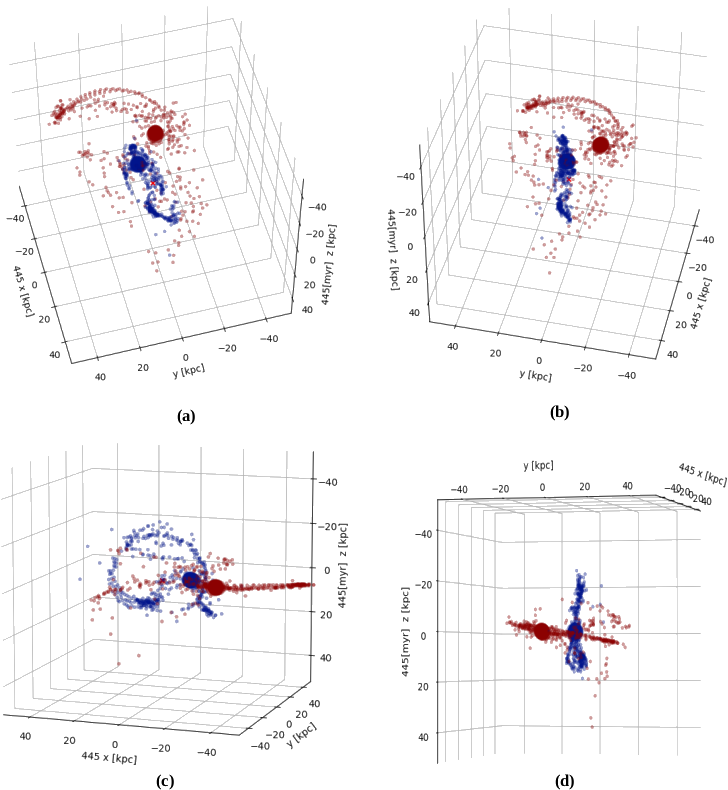}
  \caption{\label{fig:3d}Three-dimensional projection of the result of the simulation. The 3D plot has been rotated so as to view it from four different perspectives. We can see that the galaxies remain orthogonal to each other during the interaction.}
  \end{figure*}
  
To provide small alterations to the initial conditions, we rotate the galaxies around their rotation axis by small angles (five increments of $\mathrm{\pi/50}$ each), and use three different radii (radial factors of 0.97, 1.00, and 1.03) for each of these five systems. We rotate the 3D Model that we get as the result of the simulation and flatten it so that we can view it at an angle which makes the system comparable to our view of SDSS~J134420.86+663717.8. To estimate the inclinations of the galaxies in the result of the simulation, we rotate the Model by $\mathrm{90^\circ}$ around the z-axis, and then rotate it around the y-axis, such that the line of nodes is along the new line-of-sight (see Figure \ref{fig:inclination}). The rotated cube
has an orientation such that it shows the planes of the galaxies to be almost orthogonal not only to each other but also to the new line-of-sight. This allows us to roughly estimate the inclination angles by looking 
at the original line of sight and the planes of the two galaxies. They are approximately 10$\mathrm{^\circ}$ for the red galaxy and approximately 90$\mathrm{^\circ}$ for the blue galaxy.
\\
\\

\subsubsection{The predicted velocity field}\label{vel_field}
Figure \ref{fig:2d}a shows the two-dimensional projection of the Model and 
Figure \ref{fig:2d}b shows the SDSS image of the galaxies.
The comparison shows good similarities between them. The blue-coloured tail resembles the foreshortened tail associated with galaxy A to the south-west in the SDSS image, while the red-coloured tail looks like the arching tidal tail associated with galaxy B to the north-east in the SDSS image. 
The blue tail loops around the nucleus of the blue galaxy A in the model. The red tail seems to be foreshortened. Moreover, the distance between the nuclei also seems to be similar for both of the images - the simulation result and the data. 
Additionally, in Figure \ref{fig:2d}c we show the radial velocities predicted by our model. The north-western tidal arm seems to have a rather uniform velocity distribution of around -60 $\mathrm{km~s^{-1}}$, while the region where the tail meets the galaxy B body has velocity of -40 $\mathrm{km~s^{-1}}$. The velocity rises in the main body of galaxy B, such that the nuclear region of galaxy B appears to have a velocity of around 180 $\mathrm{km~s^{-1}}$ as expected from the line-of-sight velocity plot in Figure \ref{fig:rot_curve}. 
From Figure \ref{fig:2d}a, we see that the southern part of the structure has a mixture of particles from both of the galaxies, such that the (blue) particles belonging to galaxy A are enveloped on the east and the west by the (red) particles belonging to galaxy B. This is reflected in the velocity map in Figure \ref{fig:2d}c, as well. 
We see an envelope of approximately $-$40 $\mathrm{km~s^{-1}}$ to the east and $+$40 $\mathrm{km~s^{-1}}$ to west of the central structure, which shows decreasing velocities going down to -240 $\mathrm{km~s^{-1}}$ to the south-west of nucleus A. The velocity distribution along the central part of the galaxy A structure shows a lot of variation, going from -240 $\mathrm{km~s^{-1}}$ at the south-west to 240 $\mathrm{km~s^{-1}}$ in the middle to 15 $\mathrm{km~s^{-1}}$in the nuclear region. The velocity of the nuclear region for galaxy A is around 15 $\mathrm{km~s^{-1}}$. The velocities of the nuclear regions obtained from the model agree with the values estimated from the rotation curve in Figure \ref{fig:rot_curve}.

\subsubsection{Three dimensional projection}
Figure \ref{fig:3d} shows the three-dimensional projection of the simulated data. The 3D cube has been rotated so that the difference in the morphology of the galaxies based on the viewing angle is apparent. In Figure~\ref{fig:3d}a we show, for completeness, the scenario as seen by the observer in the sky. Galaxy A (blue) to the south-west and galaxy B (red) to the north-east. In Figures~\ref{fig:3d}(b-d), we show, successively, the view from directions in which either one or both the galaxies are seen edge on. We note that the galaxies remain orthogonal to each other even while interacting.

From Figure \ref{fig:3d}c, we see the blue-coloured system A almost face on.
It is evident that the distribution of matter is highly
asymmetric, and that the nucleus and a larger mass agglomeration
lie almost on opposite sides of a ring-like structure.
This corresponds to the images in Figure \ref{fig:2d}, and in particular in the
velocity map in Figure \ref{fig:2d}c, to the nuclear region of galaxy A and to the
very blue-shifted region of that system south of the nucleus A 
that coincides with the
mass agglomeration in the south visible in the top and middle
image in Figure \ref{fig:2d}a and  Figure \ref{fig:2d}b.
For the red-coloured system B we also find a highly asymmetric 
distribution of matter.
This system can be seen almost face on in Figure \ref{fig:3d}b.
Here, we have the well pronounced tidal arm to the north and the
nuclear region of the B system that is attached to it. 
At larger distances from the
nucleus B, we find highly disturbed matter associations that
are offset in velocity from the tidal arm and the nucleus, and that
have partially been lifted out of the red galactic disk
(see also Figure \ref{fig:3d}d).
Hence, we can understand why in the velocity map in Figure \ref{fig:2d}c, the highly blue-shifted part of galaxy A, showing extremely negative values of velocities, is embedded in the very extended matter from the B galaxy system, showing redshifted velocities mostly between -40 $\mathrm{km~s^{-1}}$ and 40 $\mathrm{km~s^{-1}}$.

  The striking similarities between the model and the actual galaxies show that we are on the right path, and SDSS J134420.86+663717.8 could indeed be a product of an orthogonal interaction between two spiral galaxies.

 %___________________________________________________________

\section{Conclusions and summary}\label{section5}
 
We present the optical long-slit spectroscopic data of SDSS~J134420.86+663717.8, a pair of interacting galaxies. We study their optical spectra and analyse the nature of the two galaxies that form the object. Using diagnostic diagrams and based on the presence of broad Hydrogen recombination lines, we conclude that the mechanism responsible for emission in the host of nucleus A is AGN-like, with Seyfert~1 properties, whereas nucleus B appears to be star-forming in nature. There appear to be two broad components for the Hydrogen recombination lines. This hints at a structure in the broad line region. There are almost equally massive SMBH at the centres of both host galaxies, with masses of a few times $\mathrm{10^7 M_{\odot}}$. The H$\alpha$ rotation curve of galaxy A shows some asymmetry, while the rotation curve of galaxy B appears to be uniform. There is a transition phase in the H$\alpha$ rotation curve which appears to be different from the rotation curves corresponding to galaxies A and B. We attribute this to contributions from both of the galaxies. Additionally, we use a simple N-body simulation to understand the possible initial conditions of
SDSS~J134420.86+663717.8. With initial parameters of $R_p=22.5~\mathrm{kpc}$, $i_1=-85^\circ$, $\omega_1=-10^\circ$, $i_2=10^\circ$, and $\omega_2=0^\circ$, the outcome of the simulation resembles our object rather closely. We thus conclude that SDSS~J134420.86+663717.8 is an ongoing merger of two almost orthogonal spiral galaxies with approximately equal masses. The galaxies appear to remain orthogonal during the process of interaction.

With almost equally massive hosts, SDSSJ134420.86+663717.8 constitutes a major merger. Major mergers are interesting as they can induce bursts of star-formation and feed the central SMBH, thus triggering an AGN phase (\citet{Casteels} and references therein). This appears to be true in the case of SDSSJ134420.86+663717.8. Nucleus A displays AGN behavior and has broad emission lines. \citet{Ellison2019} conclude that there is a connection between AGN triggering and merging at low $z$ (<1) for optical and mid-IR selected AGN. SDSSJ134420.86+663717.8 seems to conform to this conclusion, although, as it is a single case, it cannot be used to support or contradict the general theories pertaining to the evolution of galaxies via mergers or interaction. Additionally, it is difficult to make a general statement for all AGN due to contradictory studies (e.g., \citet{Keel, Bessiere, Satyapal, Goulding2018}, against: \citet{Gabor, Mechtley, Marian}). This requires a sample chosen based on similar criteria over the electromagnetic range.

\begin{acknowledgements}
   The authors thank the anonymous referee for their constructive comments and suggestions. \\
   Study of the conditions for star formation in nearby AGN and QSOs is carried out within the Collaborative Research Centre 956, sub-project [A02], funded by the Deutsche Forschungsgemeinschaft (DFG) – project ID 184018867. \\
   Madeleine Yttergren received financial support for this research from the International Max Planck Research School (IMPRS) for Astronomy and Astrophysics at the Universities of Bonn and Cologne. \\
   We made use of the Mods CCD Reduction package to reduce this data. ModsCCDRed was developed for the MODS1 and MODS2 instruments at the Large Binocular Telescope Observatory, which were built with major support provided by grants from the U.S. National Science Foundation's Division of Astronomical Sciences Advanced Technologies and Instrumentation (AST-9987045), the NSF/NOAO TSIP Program, and matching funds provided by the Ohio State University Office of Research and the Ohio Board of Regents. Additional support for modsCCDRed was provided by NSF Grant AST-1108693. \\
   We made use of the NASA/IPAC Extragalactic Database (NED) and of the HyperLeda database.

\end{acknowledgements}

%-------------------------------------------------------------------
\bibliographystyle{aa} 
\bibliography{references}

\begin{appendix}
\section{Estimation of the inclinations of galaxies}

  \begin{figure}[h!]
  \centering
  \includegraphics[width=0.55\textwidth]{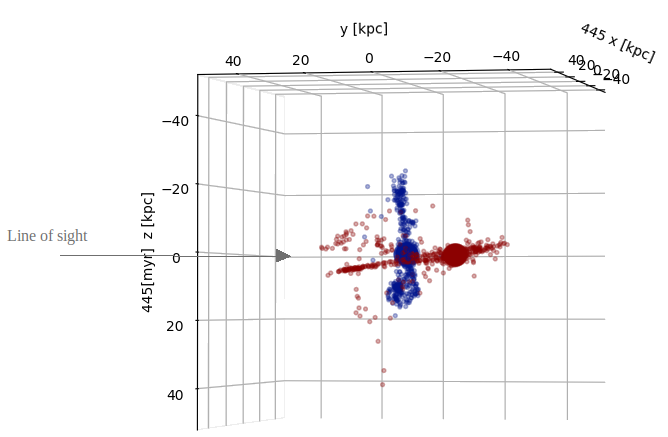}
  \caption{\label{fig:inclination} 3D model obtained as the result of the simulation is rotated by 90 degrees around the xy-plane, and then rotated about the z-axis such that the line of nodes is the new line of sight for us. The black arrow labelled 'Line of sight' shows the orientation of the plane containing the original line of sight. It can be seen that the galaxies are mutually orthogonal to each other. Additionally, the red galaxy lies along the plane of sight, while the blue galaxy is perpendicular to it.}
  \end{figure}
  
\end{appendix}

\end{document}